\documentclass[preprint,11pt]{aastex}
\usepackage[usenames]{color}
\shorttitle{3D Simulation Modeling of a Solar Flare}
\shortauthors{Nishida, Nishizuka \& Shibata}

\begin{document}
\title{The Role of a Flux Rope Ejection in Three-dimensional Magnetohydrodynamic Simulation of a Solar Flare}
\author{Keisuke Nishida\altaffilmark{1},  Naoto Nishizuka\altaffilmark{2} and Kazunari Shibata\altaffilmark{1}}
\altaffiltext{1}{Kwasan and Hida observatories, Kyoto University, Yamashina, Kyoto, 607-8471, Japan; nishida @kwasan.kyoto-u.ac.jp}
\altaffiltext{2}{Institute of Space and Astronautical Science, Japan Aerospace Exploration Agency, 3-1-1 Yoshinodai, Chuo-ku, Sagamihara, Kanagawa, 252-5210, Japan}
\slugcomment{Accepted by ApJL, August 1, 2013}

\begin{abstract}
We investigated the dynamic evolution of a 3-dimensional (3D) flux rope eruption and magnetic reconnection process in a solar flare, by simply extending 2-dimensional (2D) resistive magnetohydrodynamic simulation model of solar flares with low $\beta$ plasma to 3D model. We succeeded in reproducing a current sheet and bi-directional reconnection outflows just below the flux rope during the eruption in our 3D simulations. We calculated four cases of a strongly twisted flux rope and a weakly twisted flux rope in 2D and 3D simulations. The time evolution of a weakly twisted flux rope in 3D simulation shows similar behaviors to 2D simulation, while a strongly twisted flux rope in 3D simulation shows clearly different time evolution from 2D simulation except for the initial phase evolution. The ejection speeds of both strongly and weakly twisted flux ropes in 3D simulations are larger than 2D simulations, and the reconnection rates in 3D cases are also larger than 2D cases.  This indicates a positive feedback between the ejection speed of a flux rope and the reconnection rate even in the 3D simulation, and we conclude that the plasmoid-induced reconnection model can be applied to 3D. We also found that small scale plasmoids are formed inside a current sheet and make it turbulent. These small scale plasmoid ejections has role in locally increasing reconnection rate intermittently as observed in solar flares, coupled with a global eruption of a flux rope.
\end{abstract}


\section{Introduction}
The mechanisms of energy storage and release are remaining puzzles of solar flares \citep[see review by][]{shi11}. Magnetic reconnection converts stored magnetic energy to thermal and kinetic energies by reconnecting two anti-parallel magnetic field lines. Coronal currents store the energy required to power eruptions. Indeed, it has been shown that active regions exhibiting a sigmoidal morphology are more likely to erupt than non-sigmoidal ones \citep{can99}. It has long been suspected that solar filaments are helical in structure \citep{rus94}, and much progress has been made in modeling filament eruptions with two- and three-dimensional (2D and 3D) magnetohydrodynamic (MHD) simulations \citep{che00, ama03, tor05, kar10, kli10, kar12, kus12}.  

Magnetic shear due to slow footpoint motions in the vicinity of the polarity inversion line, especially the case of reversed shear, causes a large-scale eruption of the magnetic arcade in association with the formation of a sigmoidal structure \citep{kli10, kus12}. The eruption depends on the initial helicity, and a strongly twisted flux rope rises faster with large amount of energy release \citep{ama03}.  The initial force-free configuration with a flux rope may be linear kink unstable, when it loses the equilibrium state \citep{for94, ino06}. The helical kink instability is supposed to be one candidate for the trigger mechanism of a solar flare \citep{tor05, kar10, kli10}, as well as small scale reconnection by emerging fluxes or magnetic flux cancellation by moving magnetic features \citep{che00, moo01, ste10}.  

It is numerically shown by 2D MHD simulations that a flux rope eruption induces reconnection inflow to the current sheet and enhances both current density and electric field, finally leading to the fast reconnection \citep{che03, nisd09}. This process is named ``plasmoid-induced reconnection'' by Shibata and Tanuma (2001). Interestingly these features are verified in solar observations \citep{zhan01, qiu04, shim08, nis10} and laboratory experiments \citep{ono11}. To answer a question whether this is valid even for 3D configuration, we performed 3D resistive MHD simulation of a solar flare by simply extending 2D MHD model. In section 2, we explain our numerical model and show our results in section 3.  Finally we discuss and summarize our results in section 4.  

\section{Numerical Model}
We numerically solved the resistive MHD equations \citep[Eqs. 4--7 and 12 in ][]{shio05} in a 3D Cartesian geometry. We simply extended the previous 2D MHD model to 3D direction uniformly \citep[see 2D models;][]{che00, shio05, nisd09}. We assumed anomalous resistivity depending on current density and neglected gravity, thermal conduction and radiation cooling and heating terms. Hence, this simulation is valid only in the period when the effects of gravity, thermal conduction and radiation cooling/heating are enough small.  

\begin{figure*}[p]
\epsscale{1}
\plotone{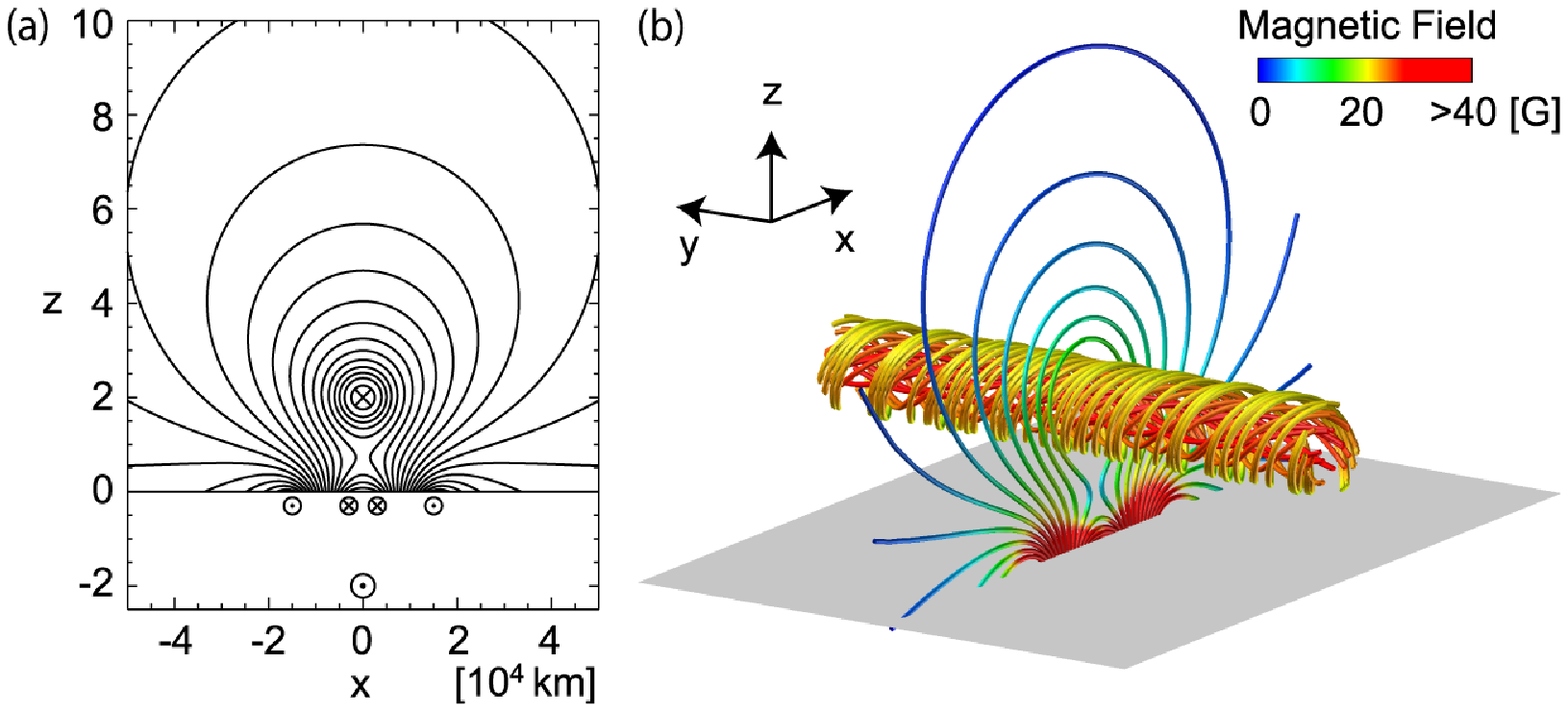}
\caption{(a) Initial magnetic field configuration of a weakly twisted flux rope in $x$-$z$ plane. Solid lines show magnetic field lines.  $\otimes$ and $\odot$ denote positive and negative line currents in $y$-direction, respectively. (b) Initial configuration of a flux rope and the ambient coronal field in 3D configuration. Color shows magnetic field strength. \label{fig1}}
\end{figure*}

Initially we assumed an isolated horizontal flux rope sustained in the corona in the almost equilibrium but unstable state. The background magnetic field is potential quadrupole field which is produced by four virtual line currents and one image current below the photosphere (Figure \ref{fig1}(a)). This configuration makes a null point above the photosphere, on which we set a flux rope with finite radius. Then we slightly changed the background configuration to keep the flux rope almost in the equilibrium state for the enough long time compared with dynamic time scale of the flux rope eruption.  

This model does not include shear (guide field) in the initial magnetic configuration. We slightly lifted up the center of the flux rope in the initial state, which triggers the loss of equilibrium state of the flux rope and then the ejection starts by the Lorentz force. This mechanism can result from an emerging flux, suddenly changing the magnetic configuration, or by the long term variation of the equilibrium state as a result of the shear motion. The gas pressure, temperature and density are initially uniform ($T_0$=10$^6$ K and $n_0$=10$^9$ cm$^{-3}$, respectively). This is unrealistic, because the realistic flux rope consists of the cool and dense plasma inside. To satisfy the force balance within the flux rope, a magnetic field, $B_y$, is added inside the flux rope. Plasma beta is $\beta \sim$0.01 at the surface of the flux rope, $\beta<$1 at lower height ($z<6L_0$) and $\beta>$1 at higher ($z>6L_0$). Here the unit length is $L_0$=10$^4$ km.

In this paper, we compared four cases: weakly and strongly twisted flux ropes in 2D and 3D simulations for comparison. The size of the computation box is (10$L_0$)$^3$ for 3D and (10$L_0$)$^2$ for 2D, and the total grid numbers in the box are 400$^3$ for 3D and 400$^2$ for 2D for comparison. The grids are uniform. Boundary condition is fixed at the bottom boundary, i.e. the whole $B$ vector is held fixed in the bottom boundary. Boundary condition is periodic in $y$-direction and free at other boundaries.

\section{Simulation Results}
\begin{figure*}[p]
\epsscale{1}
\plotone{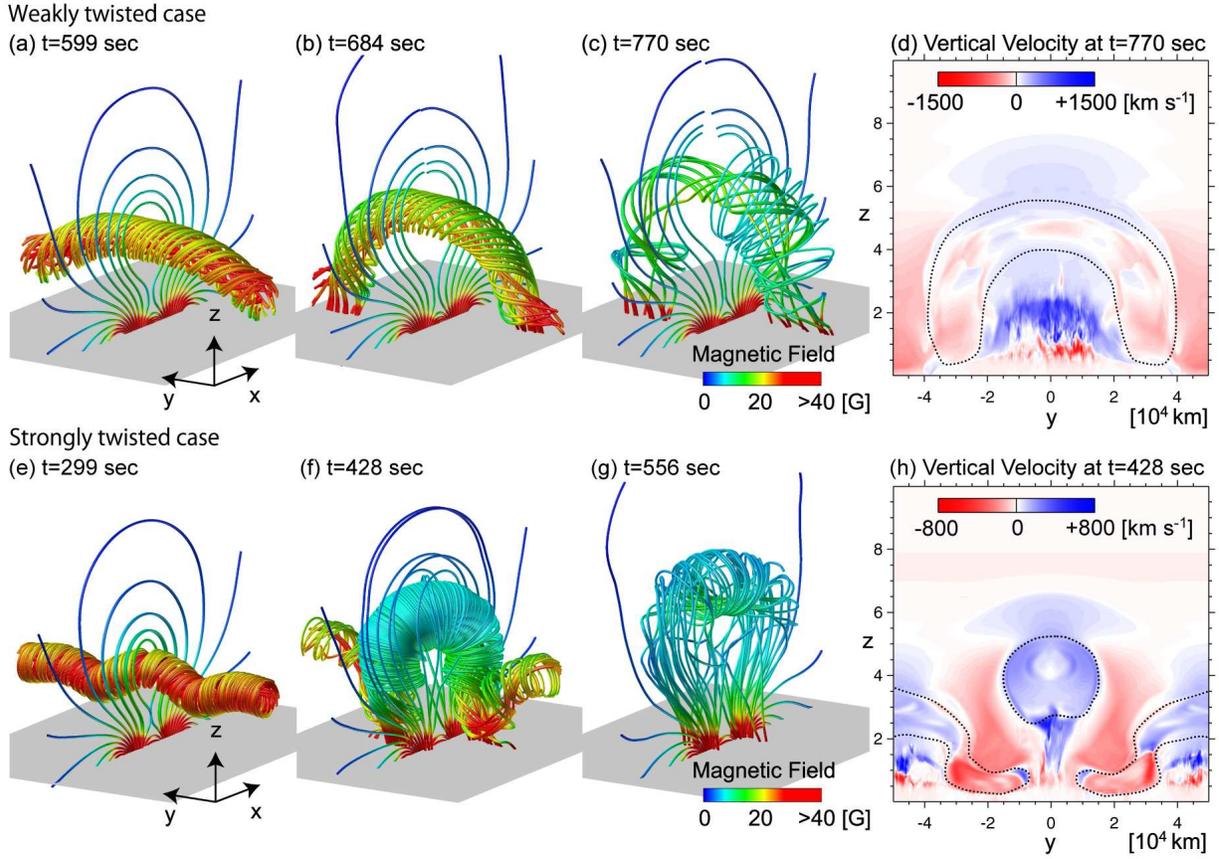}
\caption{(a)-(c) Time evolution of a weakly twisted flux rope ejection in 3D configuration. Color means magnetic field strength.  (d) Vertical velocity map in $y$-$z$ plane ($x$=0) of a weakly twisted flux rope case. Blue (red) color shows upward (downward) flow.  Dotted line shows a position of the flux rope. (e)-(g) Time evolution of a strongly twisted flux rope case. (h) Vertical velocity map in $y$-$z$ plane ($x$=0) of a strongly twisted flux rope case. \label{fig2}}
\end{figure*}

Figures \ref{fig2}(a)-\ref{fig2}(c) show snapshot images of a weakly twisted flux rope ejection in 3D simulation, in which magnetic field lines are shown with colors indicating magnetic field strength. The center of the flux rope is continuously pushed up by Lorenz force after initial slight movement, while both edges of the flux rope are forced down and collide with the bottom boundary (solar surface). As the flux rope is ejected upward, it expands because the surrounding  magnetic field decreases along the height. In the case of a weakly twisted flux rope, it moves upward keeping the 2D model like configuration, which is shown in Figure \ref{fig2}(c).

Figure \ref{fig2}(d) shows the simulation result of vertical velocity field, $v_z$ on $y$-$z$ plane ($x$=0). Blue and red colors mean upward and downward flows, respectively. Figure \ref{fig2}(d) shows intermittent bi-directional outflows at several different heights just below the flux rope. This indicates magnetic reconnection intermittently occurring along the polarity inversion line during the eruption. The maximum velocity of reconnection outflow is 1500 km s$^{-1}$ comparable to Alfv\'{e}n velocity ($V_A$). These features are consistent with observations of multiple downflows \citep{asa04, she04, mck09, sav10}. Larger number of reconnection outflows are solved with 800$^3$ grids compared with 400$^3$ grids.

For comparison, snapshot images of a strongly twisted flux rope ejection are shown in Figures \ref{fig2}(e)-\ref{fig2}(g). A strongly twisted flux rope shows initially higher upward acceleration than a weakly twisted flux rope and shows rotation about z-axis, so called writhe \citep{kli10} during the nonlinear evolution. At that time, the footpoints of the flux rope move closer \citep[similar to][]{tor05, kar10, kli10} and a current sheet becomes shorter in depth below the flux rope, while the reconnection outflows becomes smaller because of smaller reconnection magnetic field strength (Figure \ref{fig2}(h)). This is not reproduced in 2D simulation, so that this is an original 3D dynamics quite different from 2D.

The time evolutions of height and ejection speed of the two flux ropes and the electric field $\eta J$ (the reconnection rate) in 2D and 3D simulations are compared in Figure \ref{fig3}. The onset of each ejection is at $t=0$. Both strongly and weakly twisted flux ropes in 3D simulations are accelerated more strongly than in 2D simulations (Figures \ref{fig3}(b) and \ref{fig3}(e)). This is because a flux rope in 3D can be more easily ejected upward without removing or reconnecting with all the ambient magnetic field in 3D, although it must remove or reconnect with all the ambient fields in 2D. The other reason is why a flux rope is additionally accelerated by 3D effect, i.e. a force working only in 3D simulation (see Section 4). It is also stressed here that the flux rope is actually not ejected in the strongly twisted 3D case (confined or ``failed'' eruption; see Figure \ref{fig3}(d) and \ref{fig3}(e)), because magnetic tension force or the restoring force becomes stronger compared with a weakly twisted flux rope in the later phase in our configuration \citep[see also][]{tor05}. Furthermore, the reconnection speed, that is, the instantaneous peak values of the electric field $\eta J$ in the $x-z$ plane of the current sheet but averaged in $y-$direction, show the similar tendency that the reconnection speed is larger in 3D simulation rather than in 2D (Figures \ref{fig3}(c) and \ref{fig3}(f)). It is also noted here that there is a tendency that the acceleration of a flux rope associates with larger reconnection rate, though in Figure \ref{fig3}(c) the reconnection rate remains large with bursty time variation even after the upward acceleration of the flux rope level off. That is probably because small scale plasmoid ejections inside a current sheet locally and intermittently continue and increase the reconnection rate in short time periods, even when the global acceleration of a flux rope is stopped (Figures \ref{fig4} and \ref{fig5}).  

Figure \ref{fig4}  shows iso-surface of resistivity and magnetic field lines colored by magnetic field strength in the case of a weakly twisted flux rope in 3D with 800$^3$ grids. Initially a Sweet-Parker-type steady current sheet is formed below the flux rope and continues thinning until it becomes unstable for the tearing instability or the anomalous resistivity sets in. At that time, the current sheet is fragmented to several small scale current sheets with multiple X-lines and O-lines, where current density is locally enhanced. They are located at the origins of bi-directional reconnection outflows, that is, reconnection points at several different heights in Figures \ref{fig2}(d) and \ref{fig2}(h). Multi X-lines or small scale current sheets generate multiple plasmoids among themselves and eject them upward or downward intermittently, which was resolved with 800$^3$ grids simulation. This makes the current sheet turbulent and more dynamic. Simultaneously, inflows are induced to small scale current sheets, enhancing local electric field and finally driving faster reconnection (Figure \ref{fig5}). The bursty short time scale variations of the electric field $\eta J$ and the reconnection rate correspond to these small scale plasmoid ejections or the merging of these plasmoids into a single. The enhancements of the electric field may also be observed by hard X-ray emission and radio bursts (or type III burst) as a result of particle acceleration in association with harder energy release.  

\section{Summary and Discussion}
We performed 3D resistive MHD simulation of a solar flare and reproduced a flux rope eruption by simply extending 2D flare model uniformly to 3D model. After the initial slight movement of the center of the flux rope, the flux rope is ejected upward developing into the nonlinear evolution.  The ejection of a weakly twisted flux rope occurs upward keeping its time evolution in 2D $x-z$ plane quite similar to the standard 2D model. On the contrary, a strongly twisted flux rope nonlinearly evolves more rapidly, with horizontal rotation of 90 degree. The global helical shape of the erupting flux rope makes a current sheet shorter in depth below the flux rope and localizes outflow region.  

\begin{figure*}[p]
\epsscale{1}
\plotone{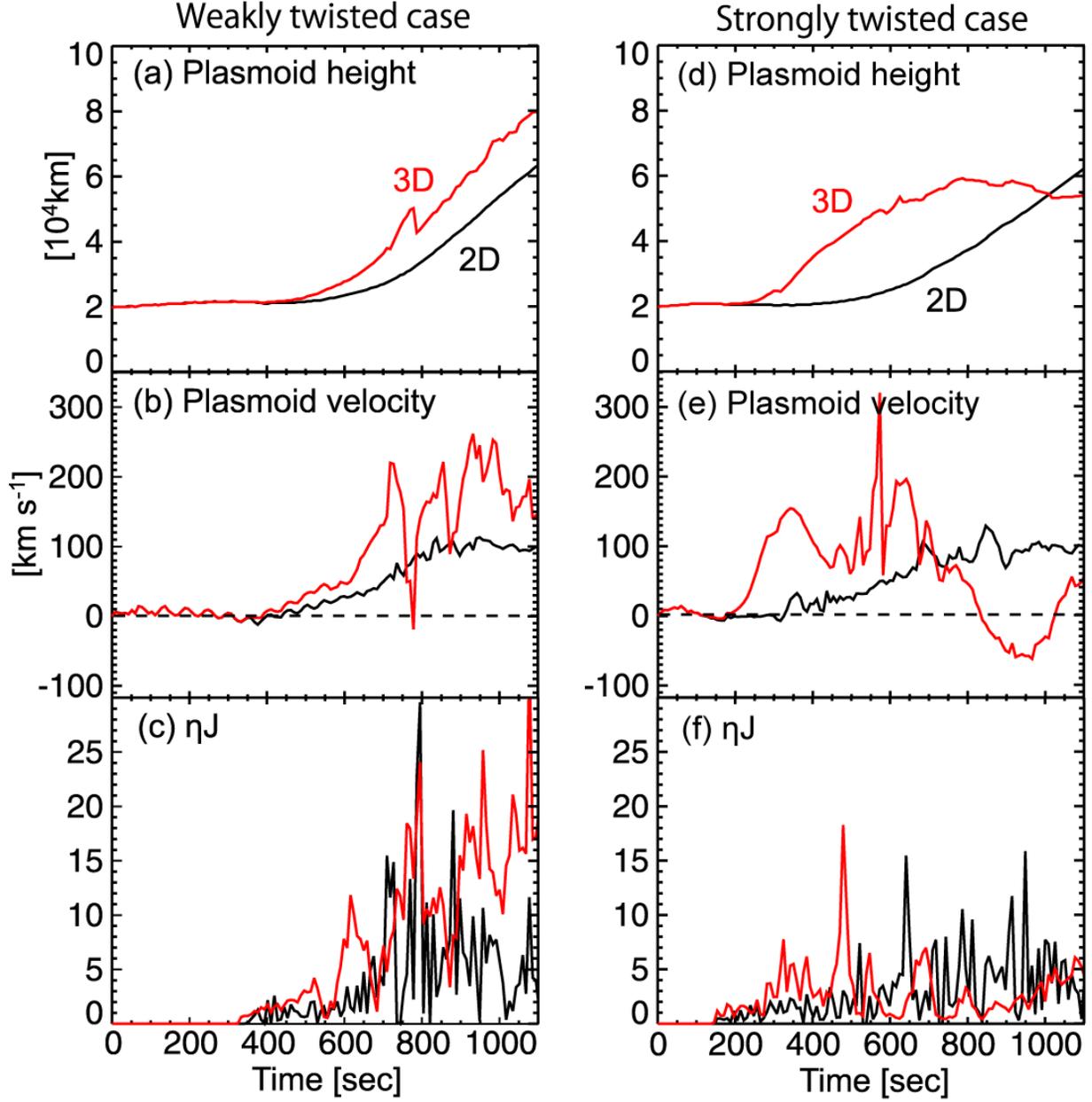}
\caption{(a) Time evolution of the height of a flux rope, (b) the ejection velocity, and (c) the electric field $\eta J$ in the reconnection region for a weakly twisted flux rope in 2D (black line) and 3D (red line) simulations. We choose peak values of electric field in each $x$-$z$ plane, and average them along $y$-direction. (d)-(f) Time evolutions of the same parameters for a strongly twisted flux rope in 2D and 3D 
simulations. \label{fig3}}
\end{figure*}

\begin{figure*}[p]
\epsscale{1}
\plotone{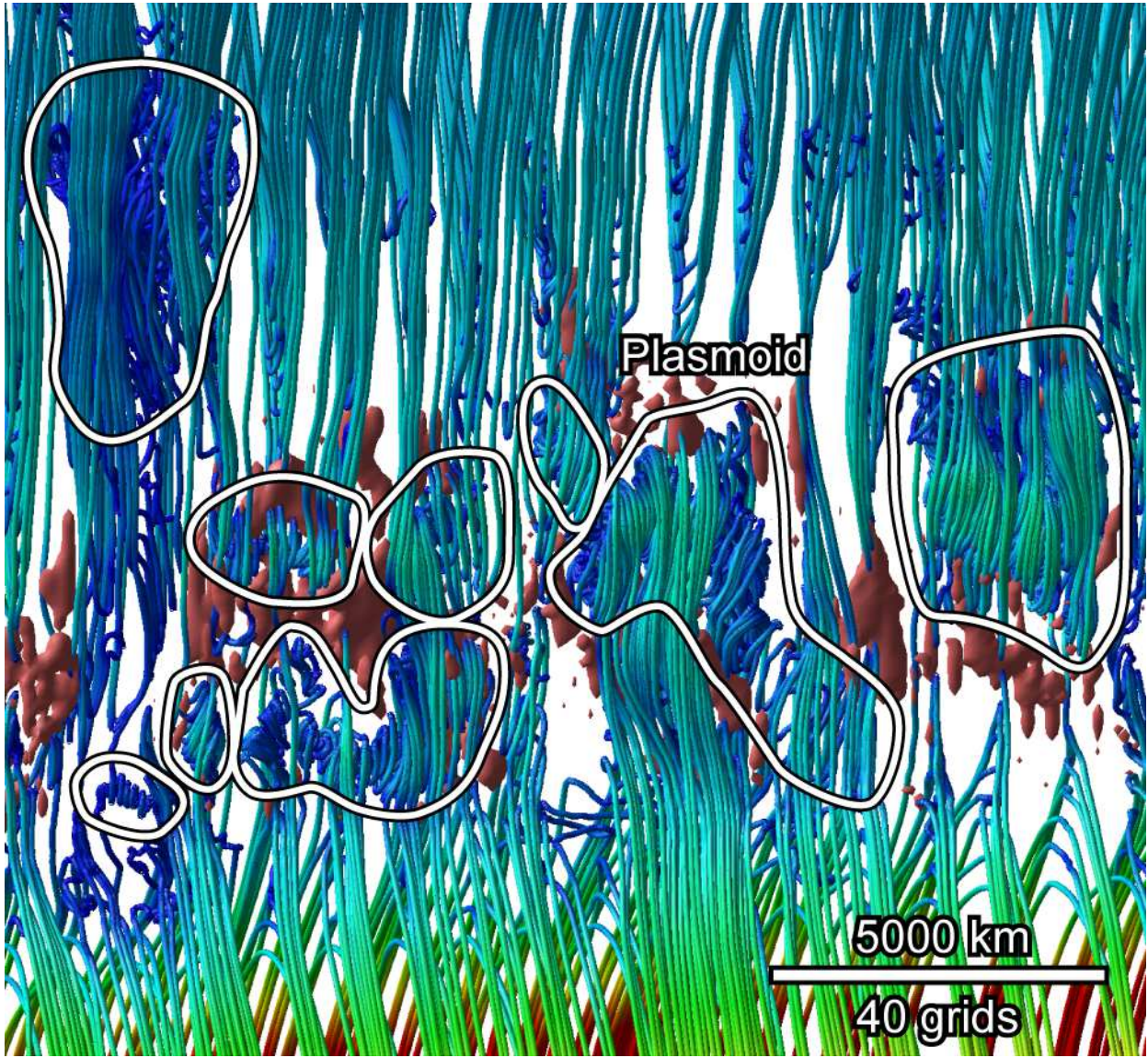}
\caption{Snapshot image of small plasmoids in a current sheet with $800^3$ grids case. Color means magnetic field strength. White lines show plasmoids. Pink surfaces show a region where the anomalous resistivity works. \label{fig4}}
\end{figure*}

\begin{figure*}[p]
\epsscale{1}
\plotone{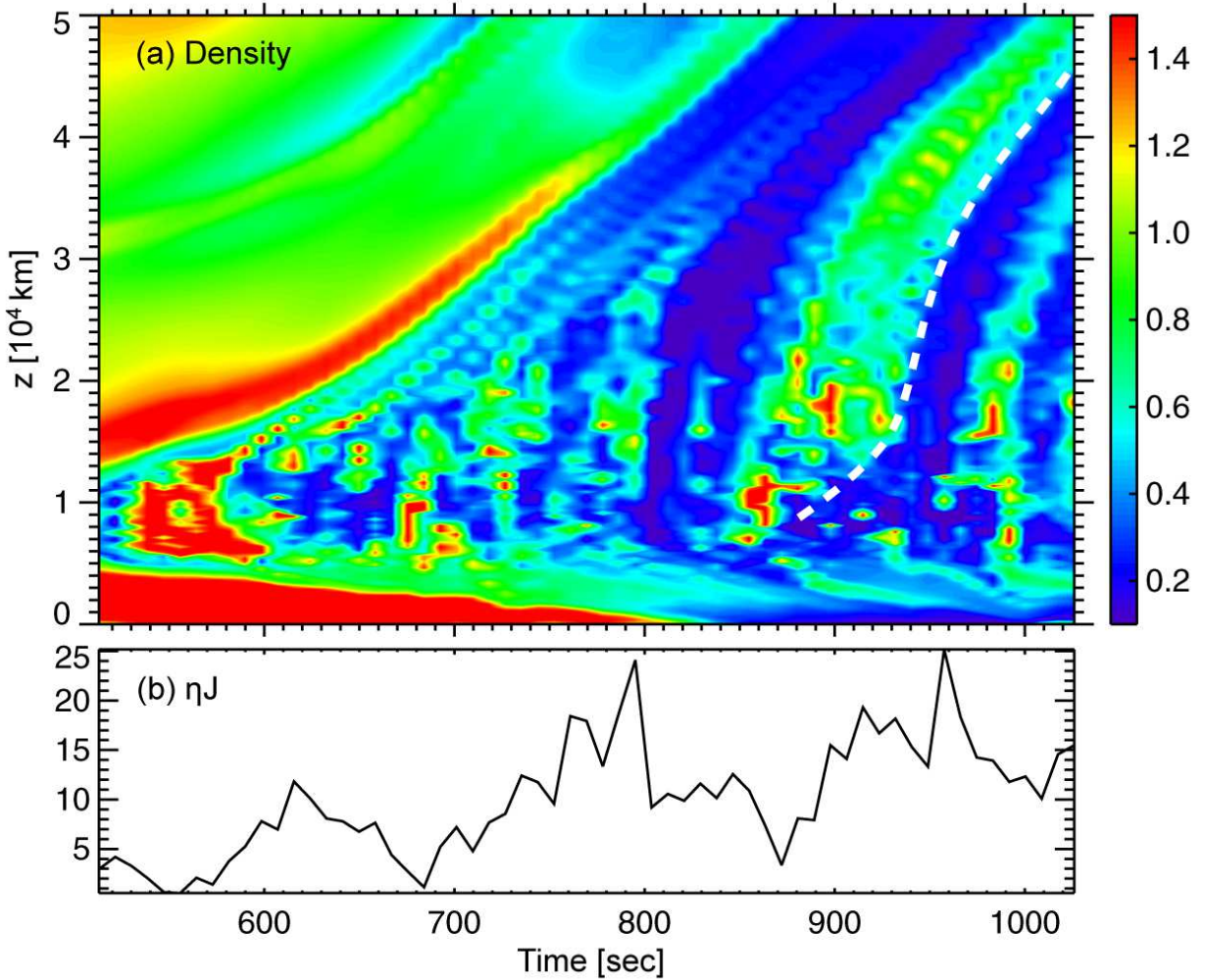}
\caption{(a) Time-height diagram of a weakly twisted flux rope at the center of the simulation box ($x=0$, $y=0$). Color means density.  White dashed line shows small plasmoid ejection in a current sheet. (b) Time evolution of the electric field $\eta J$ in the reconnection region.  We choose peak values of electric field in each $x$-$z$ plane, and average them along $y$-direction. The enhancement of electric field around $t\sim 900$ sec corresponds to the small plasmoid ejection shown as white dashed line in the panel (a).\label{fig5}}
\end{figure*}

Each flux rope in our 3D simulations reaches a higher upward velocity than the rope in the corresponding 2D simulation with the same initial configuration.  This is because a flux rope in 3D configuration can relatively easily escape from the closed coronal loops compared with 2D. This is also because the additional force by 3D effect works on a flux rope for acceleration. A flux rope with stronger twist shows larger ejection speed and larger reconnection rate in the initial phase ($t<350$ sec) (Figure \ref{fig3}(f)). This means that stronger twist enables rapid acceleration of a flux rope and consequently rapid increase of reconnection rate in the earlier phase. In the later phase after $t>350$ sec, a flux rope with stronger twist is decelerated by magnetic tension force and reconnection is suppressed, while a flux rope with weaker twist continues to be ejected upward and finally reaches up to the upper boundary.  

Next we consider the 3D additional force for flux rope acceleration. It is larger with larger amount of twist in the initial phase. It is known that a twisted flux tube with free ends is unstable to the helical kink instability. The amount by which a given line is twisted in going from one end of the tube (of length $l$) to the other is given by $\Phi$($R$)= $lB_{\phi}$($R$)/$RB_z$($R$). The flux rope is stable for kink-instability under the condition that $\Phi \le 2\pi$ \citep[Kruscal-Shafranov limit][]{bat77}. The effect of line-tying at the edges of a flux rope is stabilizing, and a uniform-twist force-free flux rope requires a twist ($\Phi$) larger than 2.5$\pi$ before it becomes kink unstable \citep{ein83}. The weakly twisted flux rope is simulated with the amount of twist $\Phi$(0)=3.0$\pi$, and the strongly twisted flux rope is $\Phi$(0)=4.5$\pi$. Therefore, the two cases in our simulation are kink unstable.  

The larger ejection velocity of the flux rope induces faster inflow to the current sheet satisfying mass conservation. At that time, magnetic flux is piled up and current density is increased. Once current density overcomes the threshold value, anomalous resistivity sets in and drives fast reconnection with large amount of released thermal energy. Inversely, fast reconnection also accelerates the ejection and evolves in the nonlinear instability. This was originally proposed as ``the plasmoid-induced reconnection'' in 2D model \citep{shi01}, and our simulation results show that this model can be applicable even to the 3D model. Here we stress that the instabilities (such as the kink-instability or the torus instability or loss of equilibrium) \citep[e.g.][]{aul10} are necessary for the eruption and formation of current sheet. However, they are not always enough to determine reconnection rate (or equivalently energy release rate). We propose that when the feedback from magnetic reconnection to the eruption or the instabilities working on the flux rope effectively works, harder energy release is enabled. Therefore, in our opinion, not only the instabilities but also the feedback from magnetic reconnection to the instabilities is important to understand energy release process in a solar flare.

We also reproduced small structures of a turbulent/fragmented current sheet, in which patchy reconnection \citep{asc01, lin06, gui11} or turbulent reconnection \citep{kow11, laz12} or fractal reconnection \citep{shi01, ji11, nis13, dra13} occurs at several heights. They produce small-scale multiple plasmoids inside and eject them out intermittently, as observed in solar flares \citep{nis10, tak11}. This makes the current sheet turbulent and more dynamic not only in 2D $x$-$z$ plane \citep[see also][]{sam09, hua10, bar11, jan11, kow11, she11, lou12, mei12} but also in 3D-direction \citep{edm10, bar12, dau12}. The tendency toward fragmentation and turbulence found in the present simulation may be quantitatively different if a stronger guide field (shear field) is included. The correlation between the flux rope/plasmoid acceleration and the reconnection rate is valid even for small-scale plasmoids inside a current sheet. This is why the coupling of global- and small-scale dynamics of plasmoid ejections explains the intermittent energy release, enhanced reconnection rate and particle acceleration in a solar flare.

\acknowledgments
We thank K. Uehara for developing the numerical code. We acknowledge the supports in part by a JSPS Research Grant, in part by the JSPS Core-to-Core Program 22001 and by the Grant-in-Aid for the Global COE Program ``The Next Generation of Physics, Spun from Universality and Emergence'' from the Ministry of Education, Culture, Sports, Science and Technology (MEXT) of Japan. The numerical computations ware carried out on the Plasma Simulator (Hitachi SR16000) at National Institutes for Fusion Science (NIFS), SX9 at Center for Computational Astrophysics (CfCA) of National Astronomical Observatory of Japan, NEC SX8 at Yukawa Institute of Theoretical Physics (YITP) in Kyoto University, and the KDK system at Research Institute for Sustainable Humanosphere (RISH) in Kyoto University.

\begin{thebibliography}{}
\bibitem[Amari et al.\ (2003)]{ama03} Amari, T., Luciani, J. F., Aly, J. J., et al., 2003, \apj, 585, 1073
\bibitem[Asai et al.\ (2004)]{asa04} Asai, A., Yokoyama, T., Shimojo, M. \& Shibata, K., 2004, \apj, 605, L77
\bibitem[Aschwanden (2001)]{asc01} Aschwanden, M. J., 2001, Particle acceleration and kinematics in solar flares (Kluwer Academic Publishers, 2001)
\bibitem[Aulanier et al.\ (2010)]{aul10} Aulanier, G., T\"{o}r\"{o}k, T., D\'{e}moulin, P., \& DeLuca, E. E., 2010, \apj, 708, 314 
\bibitem[B\'{a}rta et al.\ (2011)]{bar11} B\'{a}rta, M., B\"{u}chner, J., Karlick\'{y}, M. \& Sk\'{a}la, J., 2011, \apj, 737, 24
\bibitem[B\'{a}rta et al.\ (2012)]{bar12} B\'{a}rta, M., Sk\'{a}la, Karlick\'{y}, M., \& B\"{u}chner, J., 2012, Astrophys. Space. Sci. Proc. 33, 43
\bibitem[Bateman \& Peng (1977)]{bat77} Bateman, G., \& Peng, Y. -K. M., 1977, Phys. Rev. Lett, 38, 829
\bibitem[Canfield et al.\ (1999)]{can99} Canfield, R. C., Hudson, H. S., \& McKenzie, D. E. 1999, Geophys. Res. Lett., 26, 627
\bibitem[Chen \& Shibata (2000)]{che00} Chen, P. F., \& Shibata, K. 2000, \apj, 545, 524
\bibitem[Cheng et al.\ (2003)]{che03} Cheng, C. Z., Ren, Y., Choe, G. S., \& Moon, Y.-J., 2003, \apj, 596:1341
\bibitem[Daughton \& Roytershteyn (2012)]{dau12} Daughton, W., \& Roytershteyn, V., 2012, \ssr, 172, 271
\bibitem[Drake et al.\ (2013)]{dra13} Drake, J. F., Swisdak, M., \& Fermo, R., 2013, \apjl, 763, L5
\bibitem[Edmondson et al.\ (2010)]{edm10} Edmondson, J. K., Antiochos, S. K., DeVore, C. R., \& Zurbuchen, T. H., 2010, \apj, 718, 72
\bibitem[Einaudi \& van Hoven (1983)]{ein83} Einaudi, G. \& van Hoven, G., 1983, Sol. Phys., 88, 163
\bibitem[Forbes \& Priest (1994)]{for94} Forbes, T. G. \& Priest, E., 1994, Sol. Phys., 150, 245
\bibitem[Guidoni \& Longcope (2011)]{gui11} Guidoni, S. E., \& Longcope, D. W., 2011, \apj, 730, 90
\bibitem[Huang \& Bhattacharjee (2010)]{hua10} Huang, Y.-M. \& Bhattacharjee, A., 2010, Phys. Plasmas 17, 062104
\bibitem[Inoue \& Kusano (2006)]{ino06} Inoue, S. \& Kusano, K., 2006, \apj, 645, 742
\bibitem[Janvier et al.\ (2011)]{jan11} Janvier, M., Kishimoto, Y., \& Li, J. Q., 2011, Phys. Rev. Lett, 107, 195001
\bibitem[Ji \& Daughton (2011)]{ji11} Ji, H., \& Daughton, W., 2011, Phys. Plasmas 18, 111207
\bibitem[Karlick\'{y} \& Kliem (2010)]{kar10} Karlick\'{y}, M., \& Kliem, B., 2010, Sol. Phys. 266, 71
\bibitem[Karpen et al.\ (2012)]{kar12} Karpen, J. T., Antiochos, S. K., \& DeVore, C. R., 2012, \apj, 760, 81
\bibitem[Kliem et al.\  (2010)]{kli10} Kliem, B., Linton, M. G., T\"{o}r\"{o}k, T. \& Karlicky, M., 2010, \solphys, 266, 91
\bibitem[Kowal et al.\ (2011)]{kow11} Kowal, G., de Gouveia Dal Pino, E. M., \& Lazarian, A., 2011, \apj, 735, 102
\bibitem[Kusano et al.\ (2012)]{kus12} Kusano, K., Bamba, Y., Yamamoto, T. T., et al., \apj, 760, 31 
\bibitem[Lazarian et al.\ (2012)]{laz12} Lazarian, A., Vlahos, L., Kowal, G., et al., 2012, \ssr, 173, 557
\bibitem[Linton \& Longcope (2006)]{lin06} Linton, M. G., \& Longcope, D. W., 2006, \apj, 642, 1177
\bibitem[Loureiro et al.\ (2012)]{lou12} Loureiro, N. F., Samtaney, R., Schekochihin, A. A., \& Uzdensky, D. A., 2012, Phys. Plasmas, 19, 042303
\bibitem[Mei et al.\ (2012)]{mei12} Mei, Z., Shen, C., Wu, N., et al., 2012, \mnras, 425, 2824
\bibitem[Moore et al.\ (2001)]{moo01} Moore, R. L., Sterling, A. C., Hudson, H, S., \&  Lemen, J. R., 2001, \apj, 552, 833
\bibitem[McKenzie \& Savage\ (2009)]{mck09} McKenzie, D. E., \& Savage, S. L., 2009, \apj, 697, 1569
\bibitem[Nishida et al.\ (2009)]{nisd09} Nishida, K., Shimizu, M., Shiota, D., et al., K. 2009, \apj, 690, 748
\bibitem[Nishizuka et al.\ (2010)]{nis10} Nishizuka, N., Takasaki, H., Asai, A., \& Shibata, K., 2010, \apj, 711, 1062
\bibitem[Nishizuka \& Shibata (2013)]{nis13} Nishizuka, N \& Shibata, K, 2013, Phys. Rev. Lett., 110, 051101
\bibitem[Ono et al.\ (2011)]{ono11} Ono, Y., Hayashi, Y., Ii, T., et al., 2011, Phys. Plasmas, 18, 111213
\bibitem[Qiu et al.\ (2004)]{qiu04} Qiu, J., Wang, H., Cheng, C. Z. \& Gary, D. E., \apj, 2004, 604, 900
\bibitem[Rust \& Kumar (1994)]{rus94} Rust, D. M., \& Kumar, A. 1994, \solphys, 155, 69
\bibitem[Samtaney et al.\ (2009)]{sam09} Samtaney, R., Loureiro, N. F., Uzdensky, D. A., et al., 2009, Phys. Rev. Lett. 103, 105004
\bibitem[Savage et al.\ (2010)]{sav10} Savage, S. L., McKenzie, D. E., Reeves, K. K., et al., 2010, \apj, 722, 329 
\bibitem[Sheeley et al.\ (2004)]{she04} Sheeley, N. R., Warren, Jr., H. P., \& Wang, Y. -M., 2004, \apj, 616, 1224
\bibitem[Shen et al.\ (2011)]{she11} Shen, C., Lin, J., \& Murphy, N. A., 2011, \apj, 737, 14
\bibitem[Shibata \& Tanuma (2001)]{shi01} Shibata, K., \& Tanuma, S. 2001, Earth, Planets, and Space, 53, 473
\bibitem[Shibata \& Magara (2011)]{shi11} Shibata, K., \& Magara, T. 2011, Living Rev. Solar Phys. 8, 6
\bibitem[Shimizu et al.\ (2008)]{shim08} Shimizu, M., Nishida, K., Takasaki, H., et al., 2008, \apjl, 683, L203
\bibitem[Shiota et al.\ (2005)]{shio05} Shiota, D., Isobe, H., Chen, P. F., et al., 2005, \apj, 634, 663
\bibitem[Sterling et al.\ (2010)]{ste10} Sterling, A. C., Chifor, C., Mason, H. E., et al., 2010, A\&A, 521, A49
\bibitem[Takasao et al.\ (2011)]{tak11} Takasao, S., Asai, A., Isobe, H., \& Shibata, K., 2012, \apjl, 745, L6
\bibitem[T\"{o}r\"{o}k \& Kliem (2005)]{tor05} T\"{o}r\"{o}k, T. \& Kliem, B., 2005, \apj, 630, L97
\bibitem[Zhang et al.\ (2001)]{zhan01} Zhang, J., Dere, K. P., Howard, R. A., Kundu, M. R., \& White, S. M. 2001, \apj, 559, 452
\end{thebibliography}
\end{document}